# Learning to Reach Agreement in a Continuous Ultimatum Game

**Steven de Jong**　　　　　　　　　　　　　　　　　　　　　　STEVEN.DEJONG@MICC.UNIMAAS.NL
**Simon Uyttendaele**
*MICC, Maastricht University*
*P.O. Box 616, 6200 MD Maastricht, The Netherlands*

**Karl Tuyls**　　　　　　　　　　　　　　　　　　　　　　　　　　　　K.P.TUYLS@TUE.NL
*Eindhoven University of Technology*
*P.O. Box 513, 5600 MB Eindhoven, The Netherlands*

## Abstract

It is well-known that acting in an individually rational manner, according to the principles of classical game theory, may lead to sub-optimal solutions in a class of problems named social dilemmas. In contrast, humans generally do not have much difficulty with social dilemmas, as they are able to balance personal benefit and group benefit. As agents in multi-agent systems are regularly confronted with social dilemmas, for instance in tasks such as resource allocation, these agents may benefit from the inclusion of mechanisms thought to facilitate human fairness. Although many of such mechanisms have already been implemented in a multi-agent systems context, their application is usually limited to rather abstract social dilemmas with a discrete set of available strategies (usually two). Given that many real-world examples of social dilemmas are actually continuous in nature, we extend this previous work to more general dilemmas, in which agents operate in a continuous strategy space. The social dilemma under study here is the well-known Ultimatum Game, in which an optimal solution is achieved if agents agree on a common strategy. We investigate whether a scale-free interaction network facilitates agents to reach agreement, especially in the presence of fixed-strategy agents that represent a desired (e.g. human) outcome. Moreover, we study the influence of rewiring in the interaction network. The agents are equipped with continuous-action learning automata and play a large number of random pairwise games in order to establish a common strategy. From our experiments, we may conclude that results obtained in discrete-strategy games can be generalized to continuous-strategy games to a certain extent: a scale-free interaction network structure allows agents to achieve agreement on a common strategy, and rewiring in the interaction network greatly enhances the agents' ability to reach agreement. However, it also becomes clear that some alternative mechanisms, such as reputation and volunteering, have many subtleties involved and do not have convincing beneficial effects in the continuous case.

## 1. Introduction

Sharing limited resources with others is a common challenge for individuals in human societies as well as for agents in multi-agent systems (Chevaleyre et al., 2006). Often, there is a conflict of interest between personal benefit and group (or social) benefit. This conflict is most prominently present in a class of problems named *social dilemmas*, in which individuals need to consider not only their personal benefit, but also the effects of their choices on others, as failure to do so may lead to sub-optimal solutions. In such dilemmas, classical game theory, which assumes players or agents to be completely individually rational in strategic circumstances, seems to be of limited value (Gintis, 2001; Maynard-Smith, 1982), as individually rational players are not socially conditioned. Humans on the other hand generally show a remarkable ability to address social dilemmas, due to their tendency to consider concepts such as fairness in addition to personal benefit (see, e.g. Dannenberg et al., 2007; Fehr & Schmidt, 1999; Gintis, 2001; Oosterbeek et al., 2004).

A prime example of a social dilemma is modeled in the well-known Ultimatum Game (Gueth et al., 1982).[1] In this game, two agents bargain about the division of an amount $R$, obtained from an outsider. The first agent proposes an offer $r_2$ to the second agent (e.g. 'you receive $4 of the $10'). If the second agent accepts, each agent gets his share (i.e. the first agent receives $R - r_2$, and the second receives $r_2$); however, if the second agent rejects, both agents are left with nothing. An individually rational first agent would offer the smallest amount possible, knowing that the second agent can then choose between obtaining this amount by accepting, or nothing by rejecting. Thus, accepting the smallest amount possible is the individually rational response. In contrast, human players of the game hardly (if ever) offer less than about 20%, and if such an offer occurs, it is not likely to be accepted (Bearden, 2001; Oosterbeek et al., 2004). Thus, an individually rational player that plays as a proposer against a human player will probably not gain any money.

Researchers have proposed various mechanisms that may be responsible for the emergence of fair strategies in human populations playing social dilemmas, as well as the resistance of these strategies against invasion by individually rational strategies (see, e.g. Fehr & Schmidt, 1999; Gintis, 2001; Nowak et al., 2000; Santos et al., 2006a). Often, these mechanisms have been implemented in multi-agent systems for validation purposes, i.e. if the agents can be shown to prefer fair strategies over individually rational ones, this makes it more plausible that humans are actually affected by the underlying mechanisms. However, we argue that multi-agent systems driven by fairness mechanisms may not only be used to validate these mechanisms, but also to allow agents to act in a fair way in real-world applications. Given that agents often face tasks such as resource sharing and allocation (if not explicitly, then implicitly, e.g. sharing limited computational resources), and that these tasks regularly contain elements of social dilemmas, it is important to enable agents to act not only based on individual rationality, but also based on fairness (Chevaleyre et al., 2006). Unfortunately, existing work usually introduces a number of abstractions that do not allow the resulting multi-agent systems to be applied to realistic problems such as resource allocation. Most prominently, most work has focused on social dilemmas with discrete strategy sets (usually limited to two).[2] This abstraction simplifies the dilemmas at hand and does not reflect their potential real-world nature, since in many dilemmas, especially those related to real-world resource allocation, there is a continuum of strategies (i.e. a continuous strategy space) rather than a discrete set of pure strategies. Moreover, in social dilemmas with continuous strategy spaces, qualifications such as 'cooperation' and 'defection', which are often used in discrete social dilemmas, are actually relative: a certain strategy may be seen as cooperative (i.e. desirable) in a certain context, whereas it may be either defective or simply naive in another one. It is clear that the dilemma may be far more complicated in continuous strategy spaces, and that agents may need to use a different way of determining whether their behavior is desirable.

---

1. The analogy between the Ultimatum Game and other social dilemmas, such as the Public Goods Game, can be shown with full mathematical rigor (Sigmund et al., 2001). De Jong & Tuyls (2008) report preliminary results of applying the methodology described in this paper to the Public Goods Game.
2. Theoretical work in the field of evolutionary game theory occasionally is not limited to discrete strategy sets. Worth mentioning here is the work of Peters (2000), which introduces a theoretical extension of the evolutionary stable strategy concept (ESS) for continuous strategy spaces, i.e. the extended stability calculus. This extension provides a theoretical solution concept that can clarify egalitarian outcomes. However, the concept does not shed light on how learning agents possibly achieve these fair outcomes in social dilemmas. The work is therefore complementary to this work, which aims at mechanisms enabling agents to find fair outcomes.

In this paper, we generalize existing work on achieving agreement, cooperation and fairness in social dilemmas to continuous strategy spaces, with the aim of presenting a methodology that allows agents to reach satisfactory outcomes in these dilemmas, as well as in real-world problems containing elements of these dilemmas. We apply our proposed methodology to the Ultimatum Game, an example of a social dilemma. In this game, a population of agents needs to reach agreement (i.e. converge to a common strategy) in order to obtain a satisfactory payoff. This agreement then specifies the population's cooperative, desirable strategy. The precise nature of the agreement may vary.[3] In our population of agents, any strategy that is agreed upon by a sufficient number of agents will be successful, and will dictate the culture, or the desirable strategy, of the population. If we wish our agents to learn a 'human' strategy, we may introduce a number of simulated human agents, i.e. agents that play according to our own desirable strategy. The learning agents should then be able to imitate this strategy, even if they already reached agreement on a different, possibly relatively defective, strategy.

The remainder of this paper is structured as follows. First, in §2, we give a brief overview of the related work that this paper aims to continue. Next, in §3, we discuss our methodology, aimed at establishing agreement in large populations of learning agents. In §4, we present experiments and results. In §5, we outline a number of alternative approaches that were proposed for dilemmas with discrete strategy sets, but fail to impress in a dilemma with a continuous strategy space. We discuss why this is the case. Finally, we conclude this paper in §6.

## 2. Related Work

This work basically builds upon two tracks of existing work. We will give an overview of these tracks here and indicate how they are related to our current work. For a more extensive discussion, we refer to previous work (De Jong et al., 2008b).

### 2.1 Learning Fairness in Bargaining

De Jong et al. (2008a) investigated the behavior of agents playing the Ultimatum Game and the Nash Bargaining Game with continuous action learning automata. In both games, all agents were interacting at the same time. For the Ultimatum Game, this required an extension to more than two players in which agents, one after the other, demanded a portion of the reward at hand. The last player received what was left. The Homo Egualis utility function, as developed by Fehr & Schmidt (1999) and Gintis (2001), was used to represent a desired outcome, i.e. a required minimal amount every agent wished to obtain. Up to 100 agents were able to successfully find and maintain agreements in both games. In addition, it was observed that the solutions agreed upon corresponded to solutions agreed upon by humans, as reported in literature. In this work, we similarly use continuous action learning automata to learn agreement in the Ultimatum Game. However, our multi-agent system, organized in a network structure, can efficiently be populated with a much larger number of agents (e.g. thousands). In contrast to our previous work, agents play pairwise games. Moreover, we do not use the Homo Egualis utility function. Instead, the desired, human-inspired outcome offered by the Homo Egualis utility function is replaced here by (potentially) including agents that always play according to a certain fixed strategy (i.e. simulated human players).

---

3. Note the analogy with humans, where cultural background is one of the primary influences on what constitutes a fair, cooperative, or desirable strategy. Although there is a general tendency to deviate from pure individual rationality in favor of more socially-aware strategies, the exact implications vary greatly (Henrich et al., 2004; Oosterbeek et al., 2004; Roth et al., 1991). In the Ultimatum Game for instance, the actual amounts offered and minimally accepted vary between $10\%$ and as much as $70\%$, depending on various factors, such as the amount to bargain about (Cameron, 1999; De Jong et al., 2008c; Slonim & Roth, 1998; Sonnegard, 1996), and culture (Henrich et al., 2004). Cultural differences may persist between groups of agents, but also within these groups (Axelrod, 1997).

### 2.2 Network Topology

Santos et al. (2006b) investigated the impact of scale-free networks on resulting strategies in social dilemmas. A scale-free network was used in order to randomly determine the two agents (neighbors) that would play together in various social dilemmas, such as the Prisoner's Dilemma and the Snowdrift Game. The agents were limited to two strategies, i.e., cooperate and defect, which were initially equally probable. It was observed that, due to the scale-free network, defectors could not spread over the entire network in both games, as they do in other network structures. The authors identified that the topology of the network contributed to the observed maintained cooperation. In subsequent research, Santos et al. (2006a) introduced rewiring in the network and played many different social dilemmas, once again with two strategies per agent. They concluded that the 'ease' (measure of individuals' inertia to readjust their ties) of rewiring was increasing the rate at which cooperators efficiently wipe out defectors. In contrast to the work of Santos et al. (2006a,b), which used a discrete strategy set, this work uses a continuous strategy space. This requires another view on fairness, cooperation and agreement, departing from the traditional view that fairness is achieved by driving all agents to (manually labeled) cooperative strategies. In social dilemmas such as the Ultimatum Game, any strategy that agents may agree on leads to satisfactory outcomes.

## 3. Methodology

Before discussing our methodology in detail, we first outline our basic setting. We continue by explaining continuous-action learning automata, as they are central to our methodology. Next, we discuss the structure and topology of the networks of interaction we use. After this we discuss the agent types and initial strategies of agents. Then we elaborate on how we provide the additional possibility of rewiring connections between agents. Finally, we explain our experimental setup.

### 3.1 The Basic Setting

We study a large group of adaptive agents, driven by continuous action learning automata, playing the Ultimatum Game in pairwise interactions. Pairs are chosen according to a (scale-free) network of interaction. Every agent is randomly assigned the role of proposer or responder in the Ultimatum Game. Agents start with different strategies. For good performance, most of them need to converge to agreement by playing many pairwise games, i.e. they need to learn a common strategy. Some agents may be fixed in their strategies; these agents represent an external strategy that the adaptive agents need to converge to, for instance a preference dictated by humans. As an addition to this basic setting, we study the influence of adding the option for agents to rewire in their network of interaction as a response to an agent that behaved in a defecting manner.

### 3.2 Continuous Action Learning Automata

Continuous Action Learning Automata (CALA; Thathachar & Sastry, 2004) are learning automata developed for problems with continuous action spaces. CALA are essentially function optimizers: for every action $a$ from their continuous, one-dimensional action space $\mathbb{A}$, they receive a feedback $\beta(x)$ – the goal is to optimize this feedback. CALA have a proven convergence to (local) optima, given that the feedback function $\beta(x)$ is sufficiently smooth. The advantage of CALA over many other reinforcement learning techniques (see, e.g. Sutton & Barto, 1998), is that it is not necessary to discretize continuous action spaces, because actions are simply real numbers.

### 3.2.1 How CALA Work

Essentially, CALA maintain a Gaussian distribution from which actions are pulled. In contrast to standard learning automata, CALA require feedback on *two* actions, being the action corresponding to the mean $\mu$ of the Gaussian distribution, and the action corresponding to a sample $x$, taken from this distribution. These actions lead to a feedback $\beta(\mu)$ and $\beta(x)$, respectively, and in turn, this feedback is used to update the probability distribution's $\mu$ and $\sigma$. More precisely, the update formula for CALA can be written as:

$$\mu = \mu + \lambda \frac{\beta(x)-\beta(\mu)}{\Phi(\sigma)} \frac{x-\mu}{\Phi(\sigma)}$$
$$\sigma = \sigma + \lambda \frac{\beta(x)-\beta(\mu)}{\Phi(\sigma)} \left[\left(\frac{x-\mu}{\Phi(\sigma)}\right)^2 - 1\right] - \lambda K(\sigma - \sigma_L) \quad (1)$$

In this equation, $\lambda$ represents the learning rate; $K$ represents a large constant driving down $\sigma$. The variance $\sigma$ is kept above a threshold $\sigma_L$ to keep calculations tractable even in case of convergence.[4] This is implemented using the function:

$$\Phi(\sigma) = \max(\sigma, \sigma_L) \quad (2)$$

The intuition behind the update formula is quite straightforward (De Jong et al., 2008a). Using this update formula, CALA rather quickly converge to a (local) optimum. With multiple (e.g. $n$) learning automata, every automaton $i$ receives feedback with respect to the joint actions, respectively $\beta_i(\bar{\mu})$ and $\beta_i(\bar{x})$, with $\bar{\mu} = \mu_1, \ldots, \mu_n$ and $\bar{x} = x_1, \ldots, x_n$. In this case, there still is convergence to a (local) optimum (Thathachar & Sastry, 2004).

### 3.2.2 Modifications to CALA for our Purposes

As has been outlined above, we use CALA to enable agents to learn a sensible proposer and responder strategy in the Ultimatum Game. When playing the Ultimatum Game, two agents may agree on only one of their two joint actions (i.e. they obtain one high and one very low feedback), or may even disagree on both of them (i.e. they obtain two very low feedbacks). Both situations need additional attention, as their occurrence prevents the CALA from converging to correct solutions. To address these situations, we propose two domain-specific modifications to the update formula of the CALA (De Jong et al., 2008a).[5]

First, in case both joint actions yield a feedback of 0, the CALA are unable to draw effective conclusions, even though they may have tried a very ineffective strategy and thus should actually learn. In order to counter this problem, we introduce a 'driving force', which allows agents to update their strategy even if the feedback received is 0. This driving force is defined as:

$$\left.\begin{array}{ll}\text{For proposers:} & \beta(\mu) = \mu - x \\ \text{For responders:} & \beta(\mu) = x - \mu\end{array}\right\} \text{ iff } \beta(\mu) = \beta(x) = 0 \quad (3)$$

---

4. We used the following settings after some initial experiments: $\lambda = 0.02$, $K = 0.001$ and $\sigma_L = 10^{-7}$. The precise settings for $\lambda$ and $\sigma_L$ are not a decisive influence on the outcomes, although other values may lead to slower convergence. If $K$ is chosen to be large, as (rather vaguely) implied by Thathachar & Sastry (2004), then $\sigma$ decreases too fast, i.e. usually the CALA stop exploring before a sufficient solution has been found.
5. Note that such modifications are not uncommon in the literature; see, e.g. the work of Selten & Stoecker (1986) on learning direction theory. Grosskopf (2003) successfully applied directional learning to the setting of the Ultimatum Game, focusing on responder competition (which is not further addressed in this paper).

The effect of this modification, which we call *zero-feedback avoidance* (ZFA), is that an agent playing as the proposer will learn to offer more, and an agent playing as the responder will accept to lower his expectation. In both roles, this will lead to a more probable agreement.

Second, if one joint action yields agreement, but the other a feedback of 0, the CALA may adapt their strategies too drastically in favor of the first joint action – in fact, shifts of $\mu$ to values greater than $10^9$ were observed (De Jong & Tuyls, 2008; De Jong et al., 2008a). To tackle this problem, we restrict the difference that is possible between the two feedbacks the CALA receive in every iteration. More precisely, we empirically set:

$$\left|\frac{\beta(\mu) - \beta(x)}{\Phi(\sigma)}\right| \leq 1 \quad (4)$$

Thus, if there is a large difference in feedback between the $\mu$-action and the $x$-action, we preserve the direction indicated by this feedback, but prevent the automaton to jump too far in that direction. We call this modification *strategy update limitation* (SUL).

### 3.3 The Network of Interaction

A scale-free network (Barabasi & Albert, 1999) is a network in which the degree distribution follows a power law. More precisely, the fraction $P(k)$ of nodes in the network having $k$ connections to other nodes goes for large values of $k$ as $P(k) \sim k^{-\gamma}$. The value of the constant $\gamma$ is typically in the range $2 < \gamma < 3$. Scale-free networks are noteworthy because many empirically observed networks appear to be scale-free, including the world wide web, protein networks, citation networks, and also social networks. The mechanism of preferential attachment has been proposed to explain power law degree distributions in some networks. Preferential attachment implies that nodes prefer attaching themselves to nodes that already have a large number of neighbors, over nodes that have a small number of neighbors.

Previous research has indicated that scale-free networks contribute to the emergence of cooperation (Santos et al., 2006b). We wish to determine whether this phenomenon still occurs in continuous strategy spaces and therefore use a scale-free topology for our interaction network, using the Barabasi-Albert model. More precisely, the probability $p_i$ that a newly introduced node is connected to an existing node $i$ with degree $k_i$ is equal to:

$$p_i = \frac{k_i}{\sum_j k_j} \quad (5)$$

When we construct the network, the two first nodes are linked to each other, after which the other nodes are introduced sequentially and connected to one or more existing nodes, using $p_i$. In this way, the newly introduced node will more probably connect to a heavily linked hub than to one having only a few connections. In our simulations, we connect every new node to one, two or three existing ones (uniform probabilities). This yields networks of interaction that are more realistic than the acyclic ones obtained by always connecting new nodes to only one existing node. For example, if the network is modeling a friendship network, avoiding cycles means assuming that all friends of a certain person are never friends of each other.

### 3.4 Agent Types and Strategies

In order to study how agreement concerning a common strategy emerges, we need to make our agents learn to reach such a common strategy, starting from a situation in which it is absent (i.e. agents have different strategies). Moreover, we need to study whether a common strategy can be established from 'example' agents, and whether it is robust against agents that use a different, potentially relatively defective strategy.

#### 3.4.1 TWO TYPES OF AGENTS

We introduce two types of agents, i.e. *dynamic strategy* (DS) agents and *fixed strategy* (FS) agents. DS agents are the learning agents. They start with a certain predefined strategy and are allowed to adapt their strategy constantly, according to the learning mechanism of their learning automaton. Basically, these agents are similar to those used in earlier work (De Jong et al., 2008a). FS agents are (optional) 'good examples': they model an example strategy that needs to be learned by the (other) agents in our system, and therefore refuse to adapt this strategy.

#### 3.4.2 ONE OR TWO CALA PER AGENT?

As has been outlined above, each agent needs to be able to perform two different roles in the Ultimatum Game, i.e. playing as the proposer as well as playing as the responder. In other words, an agent is in one of two distinct states, and each state requires it to learn a different strategy. As CALA are stateless learners, each agent therefore would require two CALA. Nonetheless, in the remainder of this paper, we equip every DS agent with only one CALA, representing both the agent's proposer strategy as well as its responder strategy.

Our choice for one CALA is motivated by two observations, i.e. (1) human behavior, and (2) some initial experiments. First, human strategies are often consistent, implying that they generally accept their own offers, but reject offers that are lower (Oosterbeek et al., 2004), even with high amounts at stake (De Jong et al., 2008c; Sonnegard, 1996). Second, in a set of initial experiments, we observed that agents using two CALA will generally converge to one single strategy anyway. As an illustration, three learning curves obtained in a fully connected network of three agents playing the Ultimatum Game are displayed in Figure 1. It is clearly visible that agents' proposer strategies (bold lines) are strongly attracted to other agents' responder strategies (thin lines), and especially to the lowest of these responder strategies. In the presence of a FS agent that offers $4.5$ and accepts at least $1$, the first strategy is immediately ignored in favor of the (lower) second one. With only DS agents, once again all strategies are attracted to the lowest responder strategy present.[6]

In future work, we will study this observation from the perspective of evolutionary game theory and replicator equations (Gintis, 2001; Maynard-Smith & Price, 1973). For the current paper, we use the observation to justify an abstraction, i.e. we limit the complexity of our agents by equipping them with only one CALA. This CALA then represents the agent's proposer strategy as well as it's responder strategy. It is updated when the agent plays as a proposer as well as when it plays as a responder, according to the CALA update formula presented in §3.2.1 and the modifications presented in §3.2.2. Thus, the agents' single CALA receive twice as much feedback as two separate CALA would. This abstraction therefore increases the efficiency of the learning process.

---

6. The agents more quickly adapt their strategies downward than upward (Figure 1). Therefore, when multiple (e.g. 10) DS agents are learning (i.e. without any FS agents), their strategy usually converges to 0. This is due to an artifact of the learning process; two CALA trying to learn each others' current strategy tend to be driven downward.

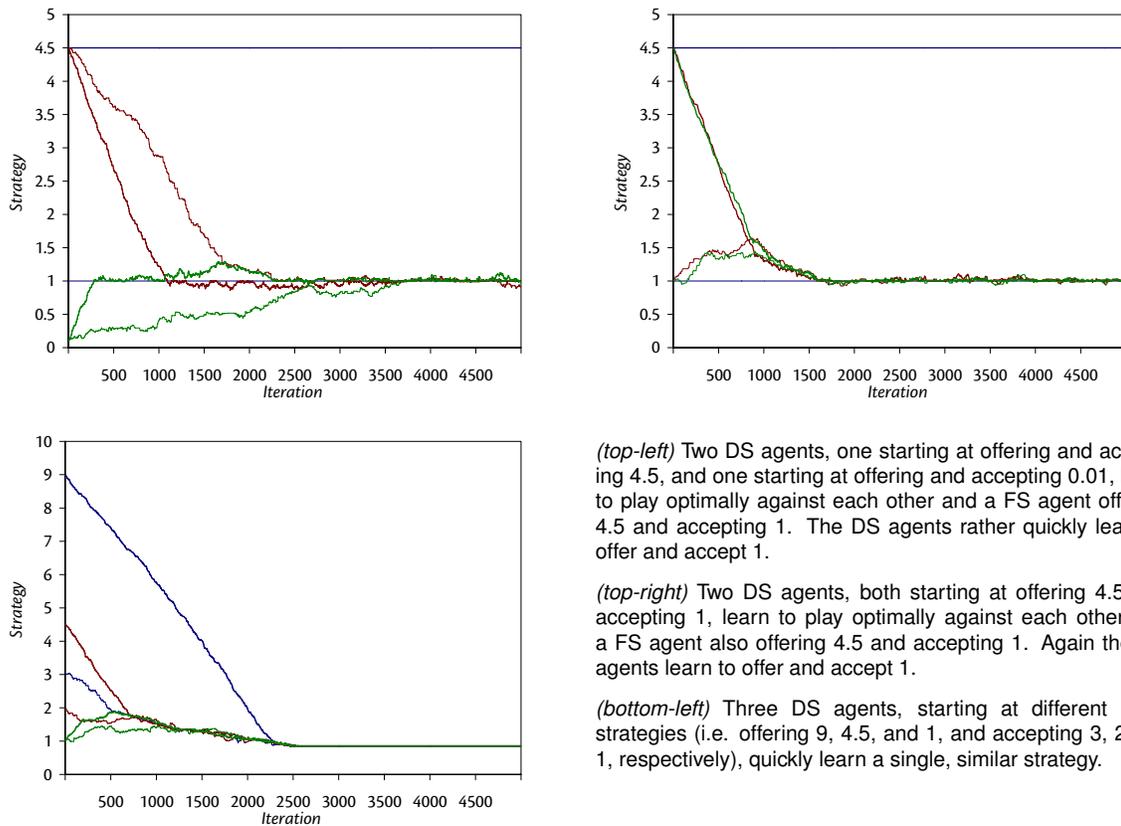

Figure 1: Evolving strategies in a fully connected network of three agents. Proposal strategies are indicated with a bold line, response strategies are indicated with a thin line. Agents converge to a situation in which their two initial strategies become similar.

### 3.4.3 AGENTS' STRATEGIES

In our simulations, we use two types of DS agents and one type of FS agents. More precisely, DSr agents are learning agents that start at a rational solution of offering $X \sim N(0.01, 1)$ (and also accepting their own amount or more). DSh agents start with a more human, fair solution, i.e. of offering $X \sim N(4.5, 1)$ (and also accepting their own amount or more). Since FS agents are examples of a desired solution, we equip them with a fair, human-inspired solution to see whether the other agents are able to adapt to this solution. The FS agents always offer $4.5$, and accept any offer of $4.5$ or more. All agents are limited to strategies taken from a continuous interval $c = [0, 10]$, where 10 is chosen as the upper bound (instead of the more common 1) because it is a common amount of money that needs to be shared in the Ultimatum Game. If any agent's strategy falls outside the interval $c$, we round off the strategy to the nearest value within the interval.

### 3.5 Rewiring

Agents play together based on their connections in the interaction network. Thus, in order to avoid playing with a certain undesirable neighbor $j$, agent $i$ may decide to break the connection between

LEARNING TO REACH AGREEMENT IN A CONTINUOUS ULTIMATUM GAME

him and $j$ and create a new link to a random neighbor of $j$ (Santos et al., 2006a).[7] For rewiring, we use a heuristic proposed by Santos et al.: agents want to disconnect themselves from (relative) defectors, as they prefer to play with relative cooperators. Thus, the probability that agent $i$ unwires from agent $j$, is calculated as:

$$p_r = \frac{s_i - s_j}{C} \quad (6)$$

Here, $s_i$ and $s_j$ are the agents' current strategies (more precisely, agent $i$'s responder strategy and agent $j$'s proposer strategy), and $C$ is the amount at stake in the Ultimatum Game, i.e. 10. Even if agents determine that they want to unwire because of this probability, they may still not be allowed to, if this breaks the last link for one of them. If unwiring takes place, agent $i$ creates a new wire to a random neighbor of agent $j$.

### 3.6 Experimental Setup

Using the aforementioned types of agents, we need to determine whether our proposed methodology possesses the traits that we would like to see. Our population can be said to have established a successful agreement if it manages to reach a common strategy that incorporates the preferences of the good examples, while at the same time discouraging those agents that try to exploit the dominant strategy. Thus, in a population consisting of only DS agents, any strategy that is shared by most (or all) agents leads to good performance, since all agents agree in all games, yielding an average payoff of 5 per game per agent – our architecture should be able to find such a common strategy. When using DS as well as FS agents, the FS agents dictate the strategy that the DS agents should converge to, regardless of whether they start as DSh or as DSr agents.

In order to measure whether the agents achieved a satisfactory outcome, we study four quantities related to the learning process and the final outcome, viz. (1) the point of convergence, (2) the learned strategy, (3) the performance and (4) the resulting network structure. We will briefly explain these four quantities below. In general, we remark that every simulation lasts for $3,000$ iterations per agent, i.e. $3,000n$ iterations for $n$ agents. We repeat every simulation 50 times to obtain reliable estimates of the quantities of interest.

#### 3.6.1 POINT OF CONVERGENCE

The most important quantity concerning the agents' learning process is the point of convergence, which, if present, tells us how many games the agents needed to play in order to establish an agreement. To determine the point of convergence, we calculate and save the average population strategy $avg(t)$ after each pairwise game (i.e. each iteration of the learning process). After $T$ iterations, we obtain an ordered set of $T$ averages, i.e. $\{avg(1), \ldots, avg(T)\}$. Initially, the average population strategy changes over time, as the agents are learning. At a certain point in time $t$, the agents stop learning, and as a result, the average population strategy $avg(t)$ does not change much anymore. To estimate this point $t$, i.e. the point of convergence, we find the lowest $t$ for which the standard deviation on the subset $\{avg(t), \ldots, avg(T)\}$ is at most $10^{-3}$. Subsequently, we report the number of games per agent played at iteration $t$, i.e. $\frac{t}{n}$. In our experiments, every simulation is repeated

---

7. Note that we may also choose to allow an agent $i$ to create a new connection to specific other agents instead of only random neighbors of their neighbor $j$. However, especially in combination with reputation (see §5.1), this allows (relative) defectors to quickly identify (relative) cooperators, with which they may then connect themselves in an attempt to exploit. Preliminary experiments have shown that this behavior leads to the interaction network losing its scale-freeness, which may seriously impair the emergence of agreement.

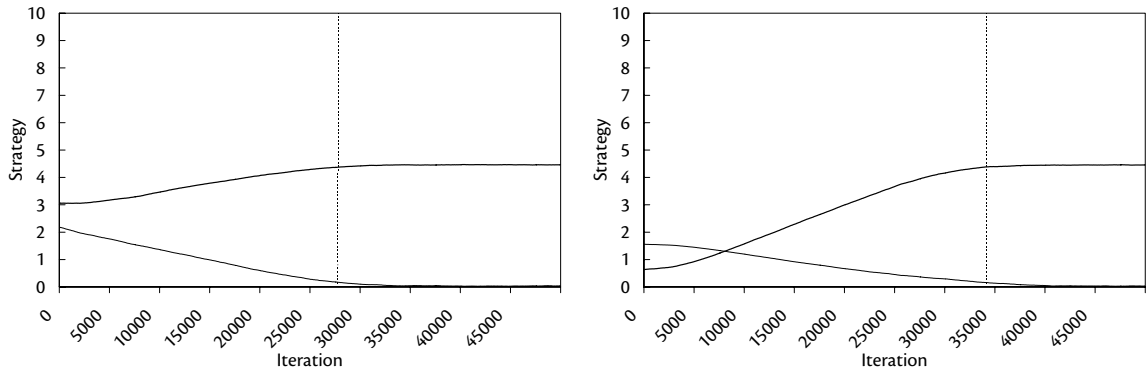

Figure 2: Two examples of the convergence point of a single run. In both graphs, we display the average strategy of the population (bold line) as well as the standard deviation on this average (thin line). The dotted vertical line denotes the convergence point, as found by the analysis detailed in the text.

50 times, resulting in 50 convergence points. We will use a box plot to visualize the distribution of these 50 convergence points.[8]

As an example, in Figure 2 (left), we see how 17 FS agents, 17 DSh agents and 16 DSr agents converge to agreement, using rewiring. Only the first $50,000$ games are shown. In addition to a bold line denoting the average population strategy, we also plot a thinner line, denoting the standard deviation on this average. Using the method outlined above, the point of convergence is determined to be around $27,500$ games, i.e. approximately 550 games per agent were necessary. In Figure 2 (right), we show similar results for 10 FS agents and 40 DSr agents, once again using rewiring. Here, the point of convergence is around $34,000$ games, i.e. approximately 680 games per agent were necessary, which means that learning to reach agreement was more difficult.

### 3.6.2 LEARNED STRATEGY

Once we established at which iteration $t$ the agents have converged, we can state that the average learned strategy is precisely $avg(t)$. We repeat every simulation 50 times to obtain a reliable estimate of this average. Once again, in our results, we use a box plot to visualize the distribution of the average learned strategy.

### 3.6.3 PERFORMANCE

To measure performance, we first allow our agents to learn from playing $3,000$ Ultimatum Games each. Then, we fix the strategies of all DS agents. We let every agent play as a proposer against all its neighbors (one by one), and count the number of games that were successful.[9] We divide this number through the total number of games played (i.e. twice the number of edges in the interaction network). The resulting number denotes the performance, which lies between 0 (for utterly catastrophic) and 1 (for complete agreement). Human players of the Ultimatum Game typically achieve a performance of 0.8–0.9 (Fehr & Schmidt, 1999; Oosterbeek et al., 2004). Once again, the 50 repetitions lead to 50 measures of performance, which are displayed in a box plot in our results.

---

8. In our box plots, we report the average instead of the median, as the average is a more informative quantity, e.g. when comparing our results with existing work. This may result in the box plots' mid point being located outside the box.
9. Note that the CALA update formula prevents agents from converging to an exact strategy, as the standard deviation of the CALA's Gaussian is kept artificially strictly positive. Therefore, there is some noise on the strategies agents have converged to. To counter this noise while measuring performance, we set responders' strategies to 99% of their actual strategies. Thus, an agent having a strategy of 4 will propose 4 and accept any offer of 3.96 or more.

### 3.6.4 RESULTING NETWORK STRUCTURE

Since the network of interaction may be rewired by agents that are not satisfied about their neighbors, we are interested in the network structure resulting from the agents' learning processes. We examine the network structure by looking at the degree distribution of the nodes in the network (i.e. the number of neighbors of the agents). With 50 repeated simulations, we may draw a single box plot expressing the average degree distribution.

## 4. Experiments and Results

We present our experiments and results in two subsections. First, we study a setup without rewiring and a setup with rewiring, varying the number of agents, while keeping the proportion of DSh, DSr and FS agents constant and equal (i.e. 33% for each type of agent). Second, we study the same two setups with various population sizes, this time varying the proportion of FS agents, where the remainder of the population is half DSh and half DSr. In general, we remark that every experiment reports results that are averaged over 50 simulations. In every simulation, we allow the agents to play $3,000n$ random games, where $n$ denotes the number of agents (i.e. the population size).

### 4.1 Varying the Population Size

In many multi-agent systems, increasing the number of agents (i.e. the population size) causes difficulties. Many mechanisms that work with a relatively low number of agents stop working well with a high number of agents, for instance due to computational complexity or undesired emergent properties. According to previous research, this issue of *scalability* also applies to the task of learning social dilemmas. Indeed, previous research using evolutionary algorithms in games with discrete strategy sets mentions that the number of games needed to converge to an agreement (i.e. on cooperation) may be "prohibitively large" (Santos et al., 2006a).[10]

Since our agents are learning in continuous strategy spaces, we may expect a scalability issue as well. To determine whether our proposed methodology has such an issue, we vary the population size between 10 and 10,000 (with some steps in between), while keeping the proportion of FS, DSh and DSr agents constant at one-third each. We study a setup without rewiring as well as a setup with rewiring, and determine (1) the point of convergence, i.e. the number of games per agent needed to reach convergence; (2) the average learned strategy the agents converged to; (3) the final performance of the system; and finally (4) the resulting network structure. Especially the first and third of these quantities give an indication of the scalability of our methodology.

---

10. In order to limit the time taken for learning, Santos et al. (2006a) terminate the learning process after $10^8$ iterations, while using at most $10^3$ agents, leading to an average of (more than) $10^5$ games per agent being available. Still, with this high number of games per agent, they report that the agents occasionally do not converge.

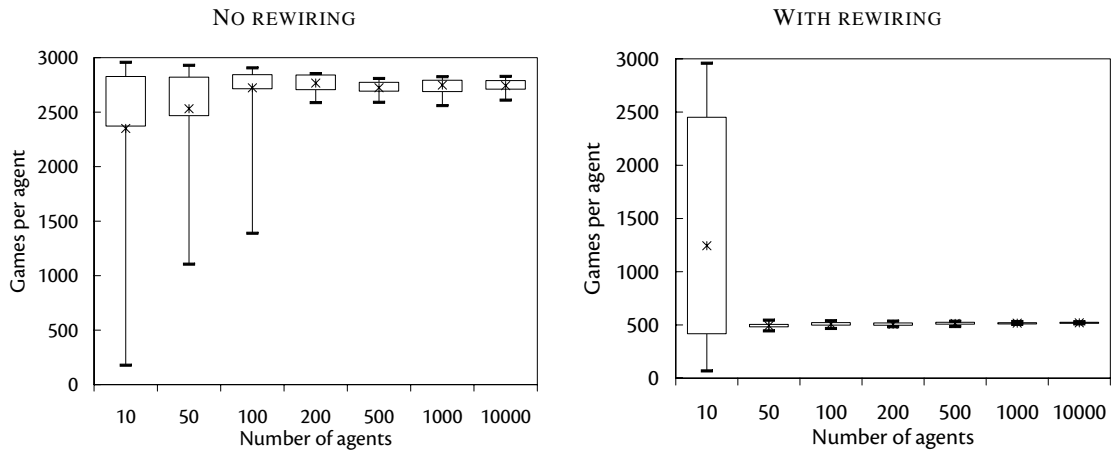

Figure 3: Games per agent until convergence; without rewiring (left) and with rewiring (right).

### 4.1.1 POINT OF CONVERGENCE

A setup without rewiring (Figure 3, left) tends to require more games per agent as the total number of agents increases. At a certain point, i.e. around a population size of 200 agents, this tendency stops, mainly because the average number of games per agent approaches the maximum, i.e. $3,000$ games per agent. A setup with rewiring (same figure, right) convincingly outperforms one without rewiring, as increasing the population size hardly affects the number of games per agent required to reach convergence. Independent from the population size, the setup requires approximately $500$ games per agent to converge. Note the difference with previous research (i.e. Santos et al., 2006a), which reports requiring $10^5$ games per agent (or more).

### 4.1.2 LEARNED STRATEGY

A setup without rewiring (Figure 4, left) on average converges to a strategy of offering as well as accepting around $3$, where $4.5$ would be required, as the $33\%$ FS agents present in the population all play with this strategy (i.e. the $66\%$ DS agents on average have a strategy of 2). With increasing population size, this average strategy is not affected; however, it becomes more and more with certainty established. Once again, a setup with rewiring (same figure, right) shows convincingly better results. Independent from the population size, the learning agents all converge to the desired strategy, i.e. $4.5$.

### 4.1.3 PERFORMANCE

With a setup without rewiring (Figure 5, left), we already saw that the average learned strategy of the DS agents is not very good. Performance is seriously affected; at around $60\%$, it indicates that few DS agents ever agree with FS agents. However, average performance is not influenced by the population size. As with the learned strategy, the performance of around $60\%$ only becomes more certainly established. As expected, a setup with rewiring (same figure, right) shows much more satisfying results, i.e. generally above $80\%$ agreement. These results are actually positively affected by the population size, as the average performance increases with an increasing population.

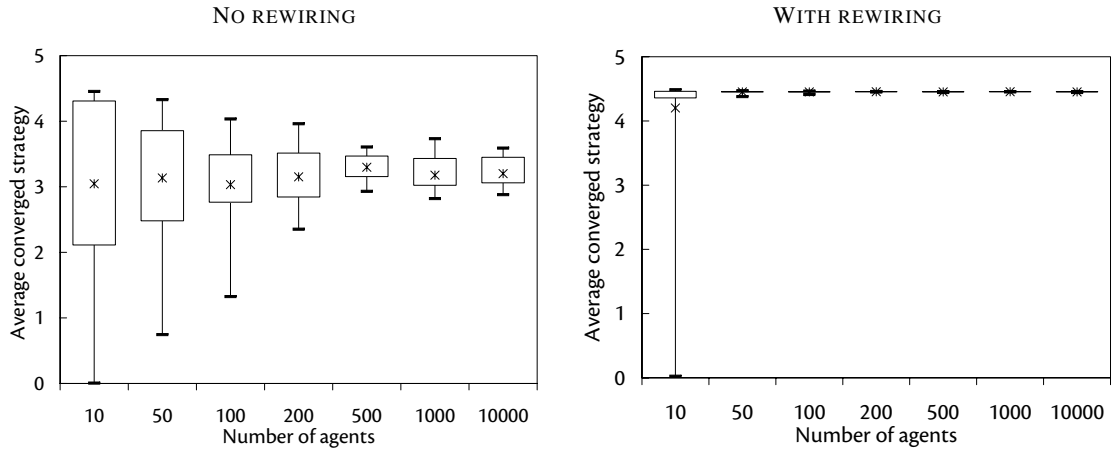

Figure 4: Average learned strategy; without rewiring (left) and with rewiring (right).

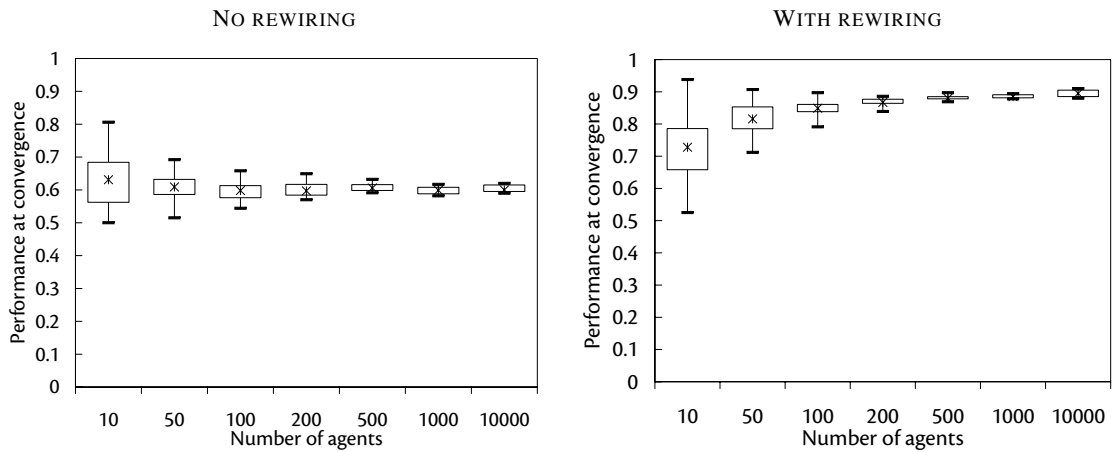

Figure 5: Final performance; without rewiring (left) and with rewiring (right).

### 4.1.4 RESULTING NETWORK STRUCTURE

We look at the network structure resulting from learning to reach agreement, and determine whether this structure is influenced by the population size. Obviously, a setup without rewiring (Figure 6, left) does not display any influence here, as the network is static. A setup with rewiring (same figure, right) shows an interesting tendency. The average degree of the resulting network stays low, while the maximum degree increases with an increasing population size. Clearly, as the population size increases, the hubs in the scale-free network receive more and more preferential attachment, and correspondingly, less densely connected nodes become even less densely connected. When we examine the number of times agents actually rewire, we find that this number generally lies below $1,000$, i.e. a very low percentage of the total number of games played actually made the agents rewire to a random neighbor of an undesired proposer.

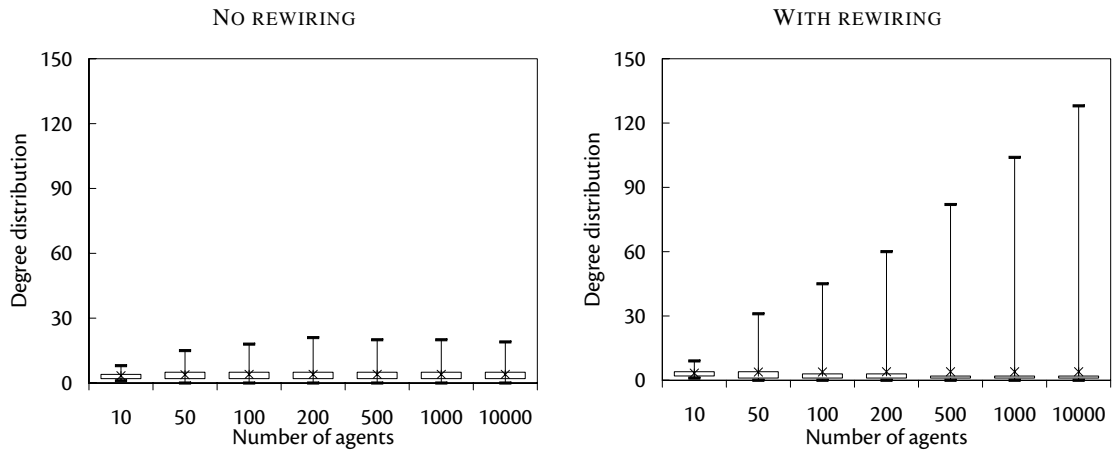

Figure 6: Resulting network structure; without rewiring (left) and with rewiring (right).

### 4.1.5 IN CONCLUSION

In conclusion to this subsection, we may state that the proposed methodology is not suffering from severe scalability issues. A setup that does not include rewiring is clearly outperformed by one that does include rewiring, but neither a setup without rewiring, nor a setup with rewiring, suffer severely from increasing the number of agents.

## 4.2 Varying the Proportion of Good Examples (FS Agents)

In this section, we investigate the behavior of the proposed methodology when the proportion of good examples in the population (i.e. FS agents with a strategy of 4.5) is varied. The remainder of the population consists of DSh and DSr agents in equal proportions. We experimented with a number of population sizes, ranging from 50 to 500.

Since the results for each population size are rather similar, we restrict ourselves to graphically reporting and analyzing the results of our experiments with 100 agents in the remainder of this section. A selection of the remaining results is given in Table 1. Specifically, for a setup without rewiring and a setup with rewiring, we report on the population size (Pop), the percentage FS agents used (%FS), the average number of games per agent needed to converge (Games), the average learned strategy (Strat), the average performance (Perf), and finally, the *maximum* number of connections that a single agent has with other agents in the network (Netw). As we will discuss below, the results reported in Table 1 for population sizes other than 100 are highly similar to those for a population size of 100 agents.

### 4.2.1 POINT OF CONVERGENCE

A setup without rewiring (Figure 7, left) requires more and more games per agent to converge, until the proportion of FS agents reaches around 30%. Then, the required number of games decreases again, although there is a great deal of uncertainty. Introducing rewiring (same figure, right) yields much better results. The number of games required per agent hardly exceeds 700, and this number decreases steadily with an increasing proportion of the population being an FS agent.

| | NO REWIRING | | | | | | WITH REWIRING | | | |
|---|---|---|---|---|---|---|---|---|---|---|
| Pop | % FS | Games | Strat | Perf | Netw | Pop | % FS | Games | Strat | Perf | Netw |
| 50 | 0 | 663.80 | 0.01 | 0.63 | 15 | 50 | 0 | 639.38 | 0.01 | 0.63 | 22 |
| | 30 | 2,588.50 | 2.87 | 0.59 | 15 | | 30 | 528.52 | 4.45 | 0.81 | 38 |
| | 50 | 1,800.02 | 4.12 | 0.70 | 16 | | 50 | 485.60 | 4.47 | 0.89 | 29 |
| | 80 | 259.86 | 4.47 | 0.87 | 15 | | 80 | 356.34 | 4.49 | 0.96 | 23 |
| 200 | 0 | 671.30 | 0.01 | 0.63 | 18 | 200 | 0 | 743.00 | 0.01 | 0.62 | 20 |
| | 30 | 2,796.85 | 2.64 | 0.57 | 17 | | 30 | 540.40 | 4.45 | 0.87 | 52 |
| | 50 | 1,354.80 | 4.17 | 0.70 | 18 | | 50 | 493.40 | 4.47 | 0.91 | 28 |
| | 80 | 288.35 | 4.47 | 0.88 | 18 | | 80 | 382.20 | 4.49 | 0.97 | 24 |
| 500 | 0 | 662.50 | 0.01 | 0.64 | 20 | 500 | 0 | 650.20 | 0.01 | 0.65 | 60 |
| | 30 | 2,793.55 | 2.85 | 0.59 | 20 | | 30 | 549.95 | 4.45 | 0.87 | 100 |
| | 50 | 1,237.75 | 4.18 | 0.69 | 21 | | 50 | 498.00 | 4.47 | 0.92 | 55 |
| | 80 | 264.60 | 4.47 | 0.89 | 21 | | 80 | 380.91 | 4.49 | 0.97 | 35 |

Table 1: Summary of the results of experiments in which the proportion of FS agents is varied. For details and additional results, see the main text.

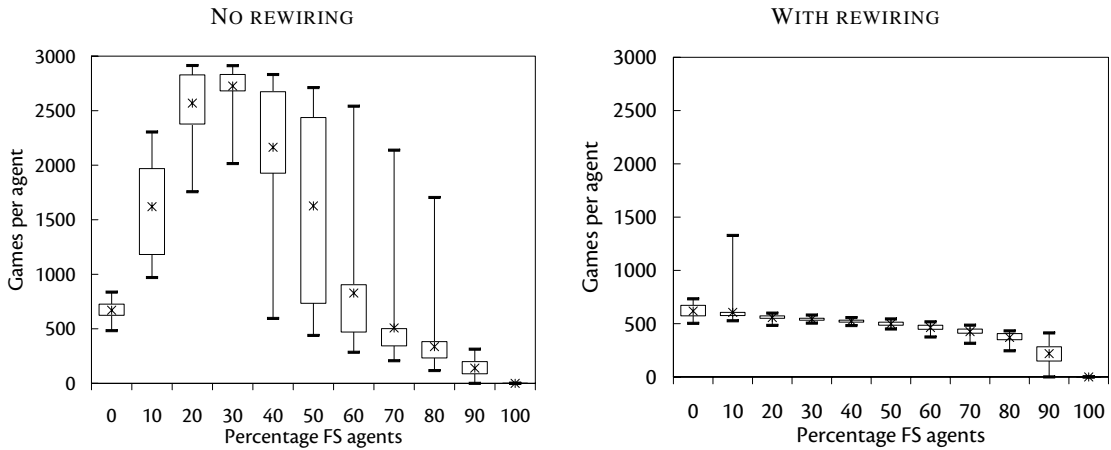

Figure 7: Games per agent until convergence; without rewiring (left) and with rewiring (right).

### 4.2.2 LEARNED STRATEGY

Interestingly, a population consisting of only DS agents tends to converge to offering and accepting the lowest amount possible, both in a setup that does not use rewiring (Figure 8, left), as well as in a setup that does (same figure, right). As has been explained in §3, DS agents tend to adapt their strategies downward more easily than upward. Thus, two DS agents that are having approximately the same strategy, may slowly pull each others' strategy downward. With many DS agents, the probability that this happens increases. Adding FS agents to the population results in different behavior for the two setups. A setup without rewiring has difficulties moving away from the lowest amount possible; only with a sufficient number of FS agents (i.e. 30% of the population) does the average learned strategy reflect that the DS agents move towards the strategy dictated by the FS agents. With rewiring, results are convincingly better; even with only 10% FS agents, the DS agents on average converge towards offering and accepting the amount dictated by these agents, i.e. 4.5.

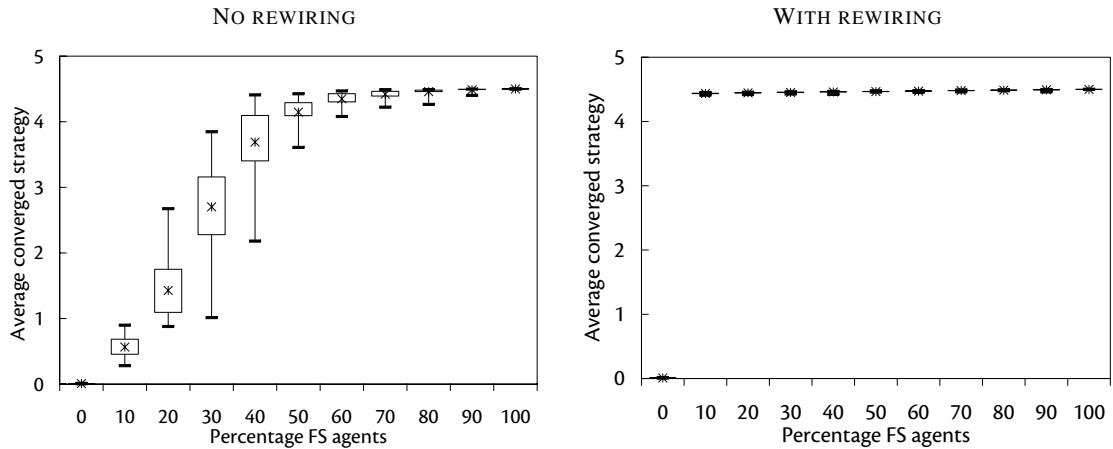

Figure 8: Average learned strategy; without rewiring (left) and with rewiring (right).

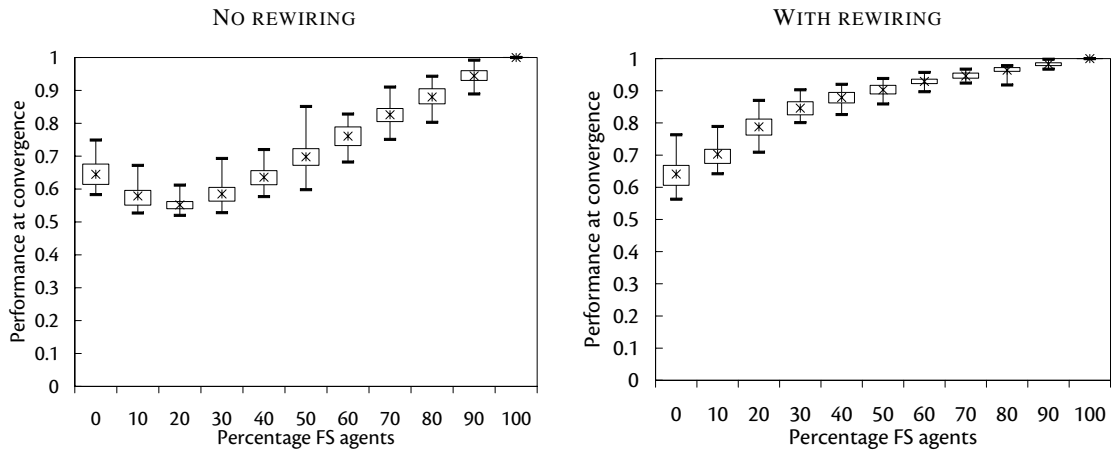

Figure 9: Final performance; without rewiring (left) and with rewiring (right).

### 4.2.3 PERFORMANCE

The observations concerning the learned strategy, as reported above, are reflected in the performance of the collective of agents. In a setup without rewiring (Figure 9, left), performance decreases initially with an increasing proportion of FS agents, as the DS agents refuse to adapt to the dictated strategy. When the proportion of FS agents becomes large enough, the DS agents start picking up this strategy, resulting in increasing performance. A setup with rewiring (same figure, right) does better, as performance increases with an increasing number of FS agents. Even though the average learned strategy is close to $4.5$ for every proportion of FS agents, low proportions of FS agents still display less performance than higher proportions. This may require additional explanation. Note that the box plot of Figure 8 shows the distribution of the average strategy over 50 repeated simulations; i.e. it does not show the strategy distribution *within* a single simulation.

Thus, even though the average strategy in a single simulation is always very close to $4.5$, there is still variance. With a low number of FS agents, this variance is most prominently caused by inertia, i.e. not all DS agents are directly connected to an FS agent, which implies that they need to learn their desired strategy from neighboring agents who are also learning. Especially with rewiring, this may result in two agents playing together that are compatible with most of their neighbors, but not (yet) with each other.

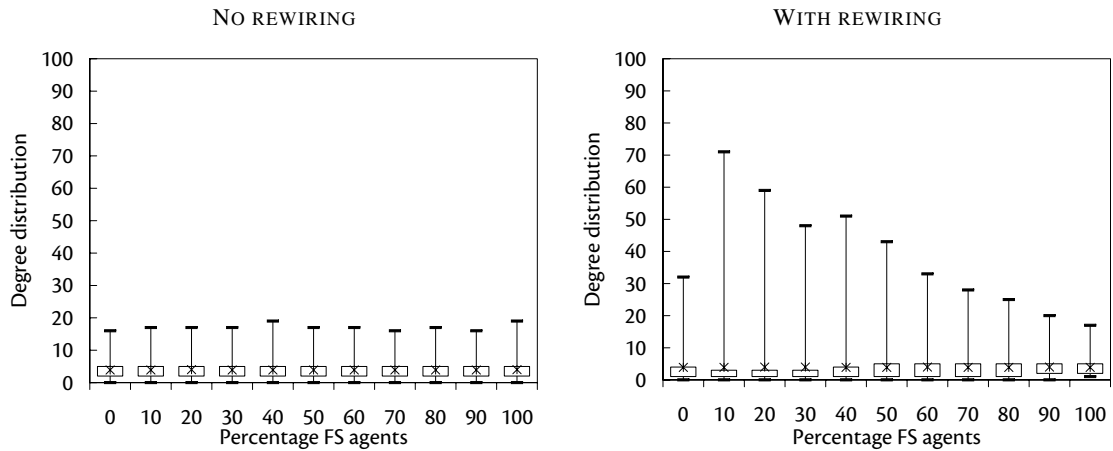

Figure 10: Resulting network structure; without rewiring (left) and with rewiring (right).

### 4.2.4 RESULTING NETWORK STRUCTURE

Clearly, the network structure of a setup without rewiring (Figure 10, left) is not influenced by varying the proportion of FS agents. When rewiring is used (same figure, right), we observe an interesting phenomenon, closely related to our observations in §4.1. Once again, the number of times agents actually rewire generally lies below $1,000$. Even though this is a low number, it does affect the network structure in a useful way. With a low proportion of FS agents, there is a large tendency for increased preferential attachment. With $10\%$ FS agents for instance, there is a single agent that connects to 70 out of 100 other agents. With an increasing proportion of FS agents, the maximum degree of the network decreases, until finally, it closely resembles the original network. Clearly, in the presence of only few examples of the desired strategy, DS agents attempt to connect to other agents that provide such examples. This is interesting and useful emergent behavior.

### 4.2.5 IN CONCLUSION

When we compare the results obtained in a population of 100 agents with the results for other population sizes, as reported in Table 1, we see that these are highly similar. In conclusion to this subsection, we may state that a setup that is not using rewiring has severe difficulties converging to a desired example if the proportion of FS agents providing this example is low. Only for, e.g. half of the population consisting of examples, does the other half learn the desired behavior. A setup that is using rewiring has absolutely no problems converging to the desired strategy, even with a low proportion of FS agents. In both cases, completely omitting the examples leads to the agents converging to the individually rational solution. This is caused by an artifact of the learning method used, i.e. as mentioned before, two CALA trying to learn each others' strategy tend to be driven downward to the lowest value allowed.

## 5. Discussion

The results presented in the previous section suggest that mechanisms that lead to cooperative solutions in social dilemmas with only a discrete set of strategies (e.g. scale-free networks and rewiring), also lead to agreement in social dilemmas with a continuous strategy space. In this section, however, we show that this is not a trivial issue. More precisely, we discuss a number of mechanisms that enhance agents' abilities to reach cooperation in social dilemmas with discrete strategy sets, but do not directly enhance agents' abilities to reach agreement in continuous strategy spaces. We empirically analyze why this is the case.

### 5.1 Reputation

Reputation is one of the main concepts used in behavioral economics to explain how fairness emerges (e.g. Bowles et al., 1997; Fehr, 2004). Basically, it is assumed that interactions between people lead to expectations concerning future interactions. These expectations may be positive or negative and may be kept to oneself, or actually shared with peers.

In work closely related to our work, Nowak et al. (2000) show that reputation deters agents from accepting low offers in the Ultimatum Game, as this information will spread, leading to the agents also receiving low offers in return. Then, if all agents refuse to accept low offers, they should provide high offers. Thus, Nowak et al. argue that the population goes toward providing and accepting high offers. However, we note that any shared strategy (i.e. any agreement) in the Ultimatum Game yields an expected payoff of $50\%$ of the amount at stake for both agents. Thus, reputation may indeed help agents to decide which strategy to play against others, but a preference for playing cooperatively (i.e. providing high offers) does not directly result from reputation.

#### 5.1.1 SPREADING REPUTATION

We study the effects of reputation by optionally adding a second network to our system. As with the interaction network, we consider the reputation network to be scale-free. In contrast to the interaction network however, the reputation network is assumed to be static, as agents are truthful concerning reputation, making it unnecessary for agents to consider rewiring. Note that two agents sharing reputation information may or may not be connected as well in the interaction network, and as a consequence, two agents playing an Ultimatum Game may or may not share reputation information with each other. In effect, after every Ultimatum Game, the responding agent may broadcast reputation information to its neighbors in the reputation network. The information is sent by the responder and concerns the offer just done by the proposer; this is the only information that is guaranteed to be correct. Agents receive information with a probability:

$$p_{ij} = 1 - \frac{d}{H} \qquad (7)$$

Here, $d$ is the distance between the sender and the (potential) receiver $j$ in the reputation network. Thus, reputation information may travel for at most $H$ hops, with a decreasing probability per hop. In our simulations, we set $H = 5$. We note that in relatively small networks, this implies that reputation information is essentially public.

Note that reputation information may be helpful only if we allow agents to *do* something with this information. In the work of Nowak et al. (2000), for instance, the reputation of others is used by agents to determine what to offer to these others. Given (1) the observation that reputation, used in this way, should not necessarily promote cooperative strategies (see above), and (2) the fact that we already use CALA to determine what agents offer to each other, we want the reputation to affect something else than agents' strategies. We will discuss a number of ways in which agents may use reputation, as taken from literature, i.e. interacting with a preferred neighbor (below) and using reputation to facilitate voluntary participation (§5.3).

### 5.1.2 USING REPUTATION

Without reputation, agents play against a random neighbor in the interaction network. Reputation may be used to make agents prefer interacting with specific neighbors – Chiang (2008) discusses that strategies of fairness could evolve to be dominant if agents are allowed to choose preferred partners to play against. Chiang allows agents to select partners that have helped the agent previously.

To determine who is a preferred partner, we use the heuristic proposed by Santos et al. (2006a), i.e. an agent prefers playing with (relative) cooperators, as these help it in obtaining a high payoff if it is the responder. Thus, the probability that agent $i$ plays with agent $j \in N_i$, where $N_i$ is the set of agent $i$'s neighbors, is:

$$p_{ij} = \frac{\tilde{s}_j - s_i}{\sum_{k \in N_i} \tilde{s}_k} \qquad (8)$$

Here, $s_i$, $\tilde{s}_j$ and $\tilde{s}_k$ are the agents' current strategies (for agents other than $i$, these are estimates based on reputation and previous experience).

There are two problems with this approach. First, the number of times that an agent $i$ receives information about an agent $j \in N_i$ may be rather low, especially with many agents. Even with only 50 agents, we observe that only around 25% of the reputation information received by agents actually concerned one of their neighbors. This problem may be addressed by making the reputation network identical to the interaction network (as neighbor relations in both networks are then identical). However, this may be seen as a considerable abstraction. Second, the probability that agent $i$ has information concerning all of his neighbors is low, so we need to specify default values for $s'_j$. Clearly, any default value is more often wrong than right, unless we use a centralized mechanism to estimate it by, for instance, using the current average population strategy, which is what we do in our simulations.

With this mechanism in place, we perform the same experiments as in §4, i.e. we vary the population size between 10 and 10,000 agents, and the proportion of FS agents in steps of 10%. A statistical analysis reveals no significant difference between a setup that uses reputation and a setup that does not. When we further analyze the results, we see that, as expected, agents almost always need to resort to default values for their neighbors' strategies. Thus, on average, the reputation system does not often change the probabilities that certain neighbors are selected.

## 5.2 Reputation and Rewiring

As we have seen in §4, rewiring works very well without reputation (i.e. purely based on an agent's own experience). Adding reputation may be beneficial to agents, as they no longer need to interact with each other to be allowed to unwire. Thus, agents may once again increase their preference for certain others. Reputation information (i.e. the amount offered by a certain agent) propagates through the (static) reputation network, allowing agents receiving such information to potentially unwire from one of their neighbors if they consider this neighbor's behavior to be undesirable. The same rewiring mechanism is used here as detailed in §3 (i.e. Equation 6). We allow the responder in the Ultimatum Game to broadcast reputation information through the reputation network, with a maximum of $H = 5$ hops.

Once again, we perform the same experiments as in §4, and once again, there is no significant difference in the main results. We further analyze the number of times agents actually rewired, and find that this number on average increases by a factor 2 with respect to a setup in which reputation is not used (i.e. as reported in §4.3). However, this increase does not increase performance. On average, agents have only few neighbors; thus, they generally receive reputation information concerning a neighbor that, in the absence of reputation, they would play against soon anyway.

### 5.3 Volunteering

According to existing research on human fairness (e.g. Boyd & Mathew, 2007; Hauert et al., 2007; Sigmund et al., 2001) the mechanism of *volunteering* may contribute to reaching cooperation in games with only two strategies. The mechanism of volunteering consists in allowing players not to participate in certain games, enabling them to fall back on a safe 'side income' that does not depend on other players' strategies. Such risk-averse optional participation can prevent exploiters from gaining the upper hand, as they are left empty-handed by more cooperative players preferring not to participate. Clearly, the 'side income' must be carefully selected, such that agents are encouraged to participate if the population is sufficiently cooperative. Experiments show that volunteering indeed allows a collective of players to spend "most of its time in a happy state" (Boyd & Mathew, 2007) in which most players are cooperative.

The biggest problem when applying volunteering is that we basically introduce yet another social dilemma. An agent may refrain from participating to make a statement against the other agent, which may convince this other agent to become more social in the future, but to make this statement, the agent must refuse an expected positive payoff: in the Ultimatum Game with randomly assigned roles, the expected payoff is always positive. Nonetheless, we study whether volunteering promotes agreement in games with continuous strategy spaces. We once again use the heuristic proposed by Santos et al. (2006a), which has already been applied in various mechanisms in this paper: if agent $i$ thinks that agent $j$ is a (relative) cooperator, then he agrees on playing. If both agents agree, then a game is played. To prevent agents from not playing any game (after all, both agents see each other as a relative cooperator only if they already are playing the same strategy), we introduce a $10\%$ probability that games are played anyway, even if one or both agents does not want to. Note that reputation may be used here, as it may allow agents to estimate whether one of their neighbors is a relative cooperator or not, without having to play with this neighbor.

Unfortunately, experimental results point out that agents using volunteering (with and without reputation) have severe difficulties establishing a common strategy (Uyttendaele, 2008). As a result, when measuring performance, we see that only around $50\%$ of the games is played. Of the games played, the performance is similar to a setup with rewiring (e.g. above $80\%$), which may be expected, as two agents usually only agree to play if their strategies are similar. The reason why agents do not converge properly is quite simple: they avoid playing with other agents that are different from them. Therefore, they do not learn to behave in a way more similar to these others.

### 5.4 General Discussion

In general, we may state that with the mechanism of rewiring, we clearly find a good balance between allowing agents to play with more preferred neighbors on the one hand, and forcing agents to learn from those different from them on the other hand. The additions discussed above allow agents to be too selective, i.e. they have too much influence on who they play against. While this may be in the interest of individual agents, it generally leads to agents not playing against others that are different from them, instead of learning from these others, as is required to obtain convergence to agreement.

## 6. Conclusion

In this paper, we argue that mechanisms thought to allow humans to find fair, satisfactory solutions to social dilemmas, may be useful for multi-agent systems, as many multi-agent systems need to address tasks that contain elements of social dilemmas, e.g. resource allocation (Chevaleyre et al., 2006). Existing work concerning (human-inspired) fairness in multi-agent systems is generally restricted to discrete strategy sets, usually with only two strategies, one of which is deemed to be 'cooperative' (i.e. desirable). However, many real-world applications of multi-agent systems pose social dilemmas which require a strategy taken from a continuous strategy space, rather than a discrete strategy set. We observed that the traditional concept of cooperation is not trivially applicable to continuous strategy spaces, especially since it is no longer feasible to manually label a certain strategy as 'cooperative' in an absolute manner. A certain strategy may be cooperative in a certain culture, whereas it may be defective or naive in another. Thus, cooperation is a relative rather than absolute concept in continuous strategy spaces.

We propose that the concept of agreement (as introduced in statistical physics; Dall'Asta et al., 2006) may be used as an alternative to cooperation. We discuss the emergence of agreement in continuous strategy spaces, using learning agents that play pairwise Ultimatum Games, based on a scale-free interaction network and the possibility to rewire in this network. In the Ultimatum Game, two agents agree if the offering agent offers at least the minimal amount that satisfies the responding agent (in this case, the two agents cooperate). Thus, for our population of agents to agree on many random pairwise games, the agents should converge to the same strategy. Without any external influences, any shared strategy is sufficient. With external influences, e.g. a preference dictated by humans, agents should adapt to the dictated strategy, even if they are already agreeing on a completely different strategy. We propose a methodology, based on continuous-action learning automata, interactions in scale-free networks, and rewiring in these networks, aimed at allowing agents to reach agreement. A set of experiments investigates the usability of this methodology.

In conclusion, we give four statements. (1) Our proposed methodology is able to establish agreement on a common strategy, especially when agents are given the option to rewire in their network of interaction. Humans playing the Ultimatum Game reach an agreement of approximately 80-90% (Oosterbeek et al., 2004). Without rewiring, our agents do worse (generally, 65% of the games are successful); with rewiring, they do as well as humans. Thus, as in games with a discrete strategy set, rewiring greatly enhances the agents' abilities to reach agreement, without compromising the scale-free network structure. This indicates that interactions in scale-free networks, as well as rewiring in these networks, are plausible mechanisms for making agents reach agreement. (2) In comparison to methodologies reported on in related work (e.g. Santos et al., 2006b), our methodology facilitates convergence with only a low number of games per agent needed (e.g. 500 instead of 10,000). This indicates that continuous-action learning automata are a satisfactory approach when we are aiming at allowing agents to learn from a relatively low number of examples. (3) The performance of the collective is not seriously influenced by its size. This is clearly influenced by the characteristics of a scale-free, self-similar network. (4) Concepts such as reputation or volunteering, which have been reported to facilitate cooperative outcomes in discrete-strategy games, do not seem to have (additional) benefits in continuous strategy spaces.

Although the Ultimatum Game is only one example of a social dilemma, its core difficulty is present in all social dilemmas: selecting an individually rational action, which should optimize one's payoff, actually may hurt this payoff. In the Ultimatum Game, this problem is caused by the fact that we may play (as a proposer) against someone who would rather go home empty-handed than accept a deal that is perceived as unfair. Similar fairness-related problems may arise in various other interactions, e.g. in bargaining about the division of a reward, or in resource allocation (Chevaleyre et al., 2006; Endriss, 2008). Many multi-agent systems need to allocate resources, if not explicitly in their assigned task, then implicitly, for instance because multiple agents share a certain, limited amount of computation time. Thus, fair division is an important area of research, which recently is receiving increasing attention from the multi-agent systems community (Endriss, 2008). As humans often display adequate and immediate ability to come up with a fair division that is accepted by most of them, it will definitely pay off if we allow agents to learn to imitate human strategies. In this paper, we examined how such a task may be executed.

## Acknowledgments

The authors wish to thank the anonymous referees for their valuable contributions. Steven de Jong is funded by the 'Breedtestrategie' programme of Maastricht University.

## References

Axelrod, R. (1997). The Dissemination of Culture: A Model with Local Convergence and Global Polarization. *Journal of Conflict Resolution*, 41:203–226.

Barabasi, A.-L. and Albert, R. (1999). Emergence of scaling in random networks. *Science*, 286:509–512.

Bearden, J. N. (2001). Ultimatum Bargaining Experiments: The State of the Art. *SSRN eLibrary*.

Bowles, S., Boyd, R., Fehr, E., and Gintis, H. (1997). Homo reciprocans: A Research Initiative on the Origins, Dimensions, and Policy Implications of Reciprocal Fairness. *Advances in Complex Systems*, 4:1–30.

Boyd, R. and Mathew, S. (2007). A Narrow Road to Cooperation. *Science*, 316:1858–1859.

Cameron, L. (1999). Raising the stakes in the ultimatum game: Evidence from Indonesia. *Journal of Economic Inquiry*, 37:47–59.

Chevaleyre, Y., Dunne, P., Endriss, U., Lang, J., Lemaître, M., Maudet, N., Padget, J., Phelps, S., Rodriguez-Aguilar, J., and Sousa, P. (2006). Issues in Multiagent Resource Allocation. *Informatica*, 30:3–31.

Chiang, Y.-S. (2008). A Path Toward Fairness: Preferential Association and the Evolution of Strategies in the Ultimatum Game. *Rationality and Society*, 20(2):173–201.

Dall'Asta, L., Baronchelli, A., Barrat, A., and Loreto, V. (2006). Agreement dynamics on small-world networks. *Europhysics Letters*, 73(6):pp. 969–975.

Dannenberg, A., Riechmann, T., Sturm, B., and Vogt, C. (2007). Inequity Aversion and Individual Behavior in Public Good Games: An Experimental Investigation. *SSRN eLibrary*.

de Jong, S. and Tuyls, K. (2008). Learning to cooperate in public-goods interactions. Presented at the EUMAS'08 Workshop, Bath, UK, December 18-19.

de Jong, S., Tuyls, K., and Verbeeck, K. (2008a). Artificial Agents Learning Human Fairness. In *Proceedings of the international joint conference on Autonomous Agents and Multi-Agent Systems (AAMAS'08)*, pages 863–870.

de Jong, S., Tuyls, K., and Verbeeck, K. (2008b). Fairness in multi-agent systems. *Knowledge Engineering Review*, 23(2):153–180.

de Jong, S., Tuyls, K., Verbeeck, K., and Roos, N. (2008c). Priority Awareness: Towards a Computational Model of Human Fairness for Multi-agent Systems. *Adaptive Agents and Multi-Agent Systems III. Adaptation and Multi-Agent Learning*, 4865:117–128.

Endriss, U. (2008). Fair Division. Tutorial at the International Conference on Autonomous Agents and Multi-Agent Systems (AAMAS).

Fehr, E. (2004). Don't lose your reputation. *Nature*, 432:499–500.

Fehr, E. and Schmidt, K. (1999). A Theory of Fairness, Competition and Cooperation. *Quarterly Journal of Economics*, 114:817–868.

Gintis, H. (2001). *Game Theory Evolving: A Problem-Centered Introduction to Modeling Strategic Interaction*. Princeton University Press, Princeton, USA.

Grosskopf, B. (2003). Reinforcement and Directional Learning in the Ultimatum Game with Responder Competition. *Experimental Economics*, 6(2):141–158.

Gueth, W., Schmittberger, R., and Schwarze, B. (1982). An Experimental Analysis of Ultimatum Bargaining. *Journal of Economic Behavior and Organization*, 3 (4):367–388.

Hauert, C., Traulsen, A., Brandt, H., Nowak, M. A., and Sigmund, K. (2007). Via freedom to coercion: the emergence of costly punishment. *Science*, 316:1905–1907.

Henrich, J., Boyd, R., Bowles, S., Camerer, C., Fehr, E., and Gintis, H. (2004). *Foundations of Human Sociality: Economic Experiments and Ethnographic Evidence from Fifteen Small-Scale Societies*. Oxford University Press, Oxford, UK.

Maynard-Smith, J. (1982). *Evolution and the Theory of Games*. Cambridge University Press.

Maynard-Smith, J. and Price, G. R. (1973). The logic of animal conflict. *Nature*, 246:15–18.

Nowak, M. A., Page, K. M., and Sigmund, K. (2000). Fairness versus reason in the Ultimatum Game. *Science*, 289:1773–1775.

DE JONG, UYTTENDAELE & TUYLS

Oosterbeek, H., Sloof, R., and van de Kuilen, G. (2004). Cultural Differences in Ultimatum Game Experiments: Evidence from a Meta-Analysis. *Experimental Economics*, 7:171–188.

Peters, R. (2000). Evolutionary Stability in the Ultimatum Game. *Group Decision and Negotiation*, 9:315–324.

Roth, A. E., Prasnikar, V., Okuno-Fujiwara, M., and Zamir, S. (1991). Bargaining and Market Behavior in Jerusalem, Ljubljana, Pittsburgh, and Tokyo: An Experimental Study. *American Economic Review*, 81(5):1068–95.

Santos, F. C., Pacheco, J. M., and Lenaerts, T. (2006a). Cooperation Prevails When Individuals Adjust Their Social Ties. *PLoS Comput. Biol.*, 2(10):1284–1291.

Santos, F. C., Pacheco, J. M., and Lenaerts, T. (2006b). Evolutionary Dynamics of Social Dilemmas in Structured Heterogeneous Populations. *Proc. Natl. Acad. Sci. USA*, 103:3490–3494.

Selten, R. and Stoecker, R. (1986). End behavior in sequences of finite Prisoner's Dilemma supergames : A learning theory approach. *Journal of Economic Behavior & Organization*, 7(1):47–70.

Sigmund, K., Hauert, C., and Nowak, M. A. (2001). Reward and punishment. *Proceedings of the National Academy of Sciences*, 98(19):10757–10762.

Slonim, R. and Roth, A. (1998). Learning in high stakes ulitmatum games: An experiment in the Slovak republic. *Econometrica*, 66:569–596.

Sonnegard, J. (1996). Determination of first movers in sequential bargaining games: An experimental study. *Journal of Economic Psychology*, 17:359–386.

Sutton, R. S. and Barto, A. G. (1998). *Reinforcement Learning: An Introduction*. MIT Press, Cambridge, MA. A Bradford Book.

Thathachar, M. A. L. and Sastry, P. S. (2004). *Networks of Learning Automata: Techniques for Online Stochastic Optimization*. Kluwer Academic Publishers, Dordrecht, the Netherlands.

Uyttendaele, S. (2008). Fairness and agreement in complex networks. Master's thesis, MICC, Maastricht University.